\documentclass[manuscript]{aastex}

\usepackage{graphicx}   % Including figure files
\usepackage{amsmath}    % Advanced maths commands
\usepackage{xcolor}     % Allow colour text

\shorttitle{Chroma+ VALD}
\shortauthors{Short}

\begin{document}


\title{The ChromaStar+ modelling suite and the VALD line list}


\author{C. Ian Short}
\affil{Department of Astronomy \& Physics and Institute for Computational Astrophysics, Saint Mary's University,
    Halifax, NS, Canada, B3H 3C3}
\email{ian.short@smu.ca}


%\altaffiltext{1}{Visiting Astronomer, Cerro Tololo Inter-American Observatory.
%CTIO is operated by AURA, Inc.\ under contract to the National Science
%\altaffiltext{2}{Society of Fellows, Harvard University.}
%\altaffiltext{3}{present address: Center for Astrophysics,
%    60 Garden Street, Cambridge, MA 02138}
%\altaffiltext{4}{Visiting Programmer, Space Telescope Science Institute}
%\altaffiltext{5}{Patron, Alonso's Bar and Grill}


\begin{abstract}

We present Version 2023-02-04 (ISO) of the Chroma+ atmospheric, spectrum, and transit light-curve
modelling suite, which incorporates the VALD atomic line list.  This is a major
improvement as the previous versions used the much smaller NIST line list.  The NIST line
list is still available in Chroma+ for those projects requiring speed over completeness of
line opacity.
We describe a procedure for exploiting the ''Array job'' capability
of the slurm workload manager on multi-cpu machines to compute broadband high resolution
spectra with the VALD line list quickly using the Java version of the code (ChromaStarServer (CSS)).
The inclusion of 
a much larger line list more completely allows for the many weaker lines that over-blanket the
blue band in late-type stars and has allowed us to reduce the amount of additional
{\it ad hoc} continuous opacity needed to fit the solar spectral energy distribution
(SED).  The additional line opacity exposed a subtle bug in the spectrum synthesis
procedure that was causing residual blue line wing opacity to accumulate at shorter
wavelengths.  
We present our latest fits to the observed solar SED and to the observed
rectified high resolution visible band spectra of the Sun and the standard stars 
Arcturus and Vega.  We also introduce the fully automated Burke-Gaffney Observatory (BGO) 
at Saint Mary's University (SMU) and compare our synthetic spectra to low resolution 
spectra obtained with our grism spectrograph that is available to students. 
The fully automated BGO, the spectrograph, and the BGO spectrum reduction procedure
are fully described in a companion paper.
All codes are available from the OpenStars www site:
www.ap.smu.ca/OpenStars.
 
\end{abstract}


\keywords{Stars: atmospheres (1621), late-type (909)}

\section{Introduction}

 In \citet{shortb21} and papers in that series we have described an integrated cross-platform atmospheric modelling, spectrum synthesis and transit light-curve
 modelling code (the Chroma+ suite) developed in platform-independent languages including Python (ChromaStarPy (CsPY)), Java (ChromaStarServer (CSS)), and Javascript.  In particular, \citet{shortb21}
describe incorporation of the GAS package into CsPy and CSS, providing the codes with a mature,
competitive module for handling the combined chemical equilibrium, ionization equilibrium,
and equation-of-state (EOS) problem for over a hundred atomic, ionic, diatomic, and polyatomic 
species, which in turn allows for more realistic calculation of the line and continuum extinction
distributions, $\kappa_\lambda(\tau)$.
 The motivation has been to provide
 a more widely accessible numerical laboratory for rapid numerical experiments in spectrum synthesis and light-curve modelling, and a responsive
 electronic spectral atlas for quick spectral reconnaissance with {\it ad hoc} stellar parameters.  To expedite our proof-of-concept, we initially
 prepared a limited atomic line list based on the NIST atomic database \citep{nist} containing $\sim 26\, 000$ lines in the $\lambda 260$ to $2600$ nm 
 region (2.9 Mbytes).  With Version 2023-02-04, we have prepared a new, larger atomic line list based on the VALD 
 database (\citet{VALD19}, \citet{VALD15}) containing $\sim 613\, 000$ lines in the 
 $\lambda 250$ to $2600$ nm
 region (36 Mbytes) for as many as six ionization stages of all elements up to and including Ge ($z=32$) and additional select elements up to La ($z=57$).  
 Because responsiveness is one of the key distinguishing characteristics of the Chroma+ suite, CsPy and CSS provide a new
 input parameter that allows the user to select between the NIST or VALD lists.  Molecular band opacity continues to be treated in the 
 Just-Overlapping-Line-Approximation (JOLA, \citep{jola}) and we do not require a molecular line list.  

\section{Related improvements \label{improve}}

The VALD line list includes $\sim 70\, 000$ lines of \ion{Fe}{1} and $\sim 80\, 000$ lines of \ion{Fe}{2} and represents a $\sim 24\times$ 
increase in the number of potential lines considered in the CSS spectrum synthesis procedure.  This additional line
opacity allows the Chroma+ suite to more accurately model the effect of over-blanketing in the blue band of GK stars. 
As reported in \citet{short16} the Chroma+ suite uses an unusual method of sampling the $\lambda$ range over which each spectral line profile,
$\phi_\lambda(\lambda-\lambda_{\rm 0}$), is distributed about line center ($\lambda_{\rm 0}$) based on how spectral lines are treated in 
non-local thermodynamic equilibrium (NLTE) by the PHOENIX code \citep{phoenix}.  As each spectral line that passes an initial 
test of line-to-continuum opacity at line center (${\kappa^{\rm l}_{\lambda_0} \over \kappa^{\rm c}_{\lambda_0}} > \epsilon$) at
three reference $\log\tau_{\lambda_0}$ values throughout the atmosphere
is added to the extinction distribution spectrum ($\kappa_\lambda(\tau)$),
its own line-specific grid of $\lambda_0$-centered $\lambda$ points is inserted into the master $\lambda$ grid.  After all
lines have been added, the grid is swept of $\lambda$ points for which $\Delta\lambda$ is less than a $\Delta\lambda_{\rm min}$ value of
$\sim 0.001$ nm, approximately consistent with a numerical resolving power of $\sim 500\, 000$.
This has the advantage of minimizing the number of $\lambda$ points at which the $\kappa_\lambda(\tau)$ distribution and the corresponding 
monochromatic emergent intensity, $I_\lambda (\tau_\lambda = 0)$ must be computed in cases where there are under-blanketed spectral regions.
The large increase in the density of spectral lines provided by the VALD line list revealed a subtle bug in our procedure:  As each line is
added, the previous $\kappa(\lambda)$ distribution is interpolated onto the updated one, and a small residual line opacity at the
first $\lambda$-point in each line-specific grid was accumulating at all $\lambda$ values of $\lambda < \lambda_0$ each time a new line was added,
eventually amounting to significant spurious excess continuum opacity progressively at smaller $\lambda$ values.  This bug has now been fixed.
We also improved our $\lambda$ interpolation procedure to ensure a symmetric distribution of $\lambda$ points about the $\lambda_0$ value
of all lines more reliably with the result that broad lines now have wings that appear more symmetric.

\paragraph{}

\begin{figure}
\includegraphics[width=\columnwidth]{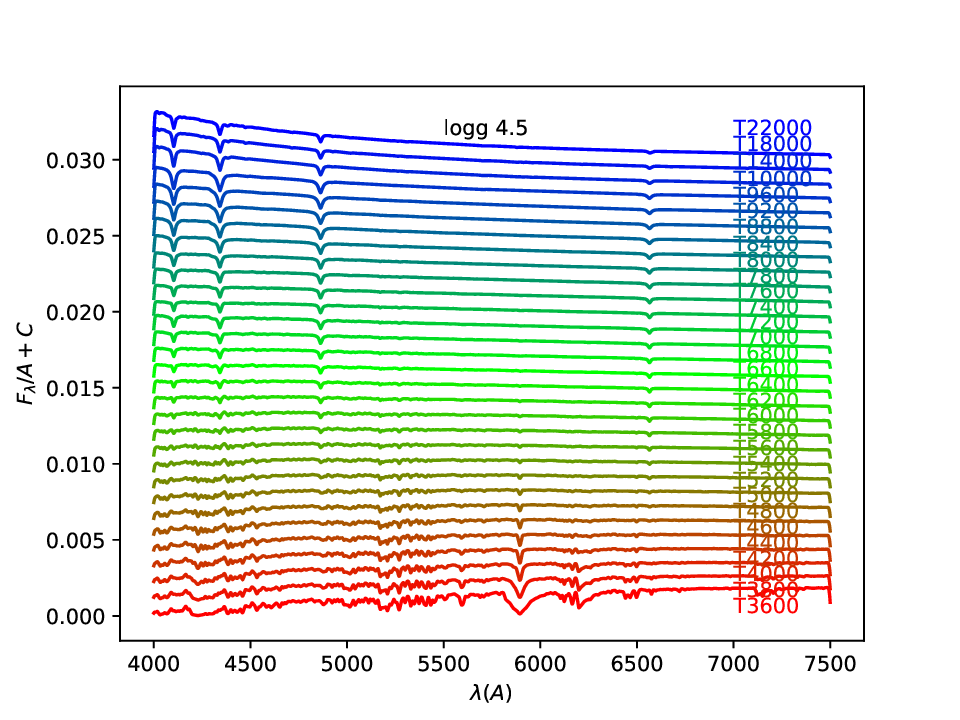}
\caption{Sequence of 31 synthetic spectra computed with CSS and the VALD line list for 
$\log g=4.5$, $[{{\rm A}\over{\rm H}}]=0.0$
and $T_{\rm eff}$ values in the range 3600 to 22\, 000 K with 
$\Delta T_{\rm eff}$ intervals of 200 K for $T_{\rm eff} \le 8000$ K and 400 K for 
$T_{\rm eff} > 8000$ K.
  \label{showspec}
}
\end{figure}

We have used the more complete spectrum synthesis to perform the following, in order: 1) Reduce the amount of {\it ad hoc} extra 
continuous opacity needed throughout the visible band (''opacity fudge'') to fit the overall observed SED of the Sun with a model of 
solar input parameters from a factor of $\sim 3$ to a factor of $\sim 1.5$., and 2) Re-calibrate the fine tuning of the 
linear Stark broadening of \ion{H}{1} Balmer lines in stars of
spectral class B to F V.  
Fig. \ref{showspec}
shows the synthetic spectra in the $\lambda$ 400 to 700 nm region for 31 models spanning the
$T_{\rm eff}$ range 3600 to 22\, 000 K with $\log g = 4.5$ and $[{{\rm A}\over{\rm H}}] = 0.0$,
with $\Delta T_{\rm eff}$ intervals of 200 K for $T_{\rm eff} \le 8000$ K and 400 K for
$T_{\rm eff} > 8000$ K.  We have also computed structures and spectra for models of $\log g = 2.0$ 
at select $T_{\rm eff}$ values less than 5000 K.

\section{Performance \label{perf}}

The Slurm workload manager provided on the Digital Research Alliance of Canada (DRAC) parallel computing clusters allows for ''Array'' jobs in which
parts of the execution that can be automatically associated with a CPU identifier are automatically scattered to the corresponding CPUs.
The spectrum synthesis calculations at a given $\lambda$ value are independent of those at another, and we use the Array job capability to 
automatically parallelize the wavelength domain of the spectrum synthesis.  
Because the density of spectral lines
per unit $\Delta\lambda$ interval increases with decreasing $\lambda$ value, we distributed the splice points in the wavelength 
parallelization in approximately equal $\Delta\log\lambda$ intervals, finessed {\it ad hoc} to avoid splice points falling in the wings of lines 
that are broad in stars of the input stellar parameter values.  
%Fig. \ref{SunSed} shows the location of the splice points for the case of the
%Sun's spectrum.  
For our performance test, we divided the $\lambda 400$ to $750$ nm range into 
eight logarithmic sub-ranges, each of which was handled by one of eight CPUs.  Because spectrum synthesis is by far the slowest major stage in 
our integrated modeling process, we achieved a significant reduction in wall-clock time.  The slurm ''sacct'' and ''seff'' reporting tools 
indicate CPU times and memory allocations ranging from 22 minutes and 0.7 Gbytes to 80 minutes and 1.2 Gbytes with both values generally 
increasing with decreasing $\lambda$ values of the assigned $\lambda$ range.  

%\begin{equation}
\section{Comparison to standard stars} 

\subsection{The Sun} 

\begin{figure}
\includegraphics[width=\columnwidth]{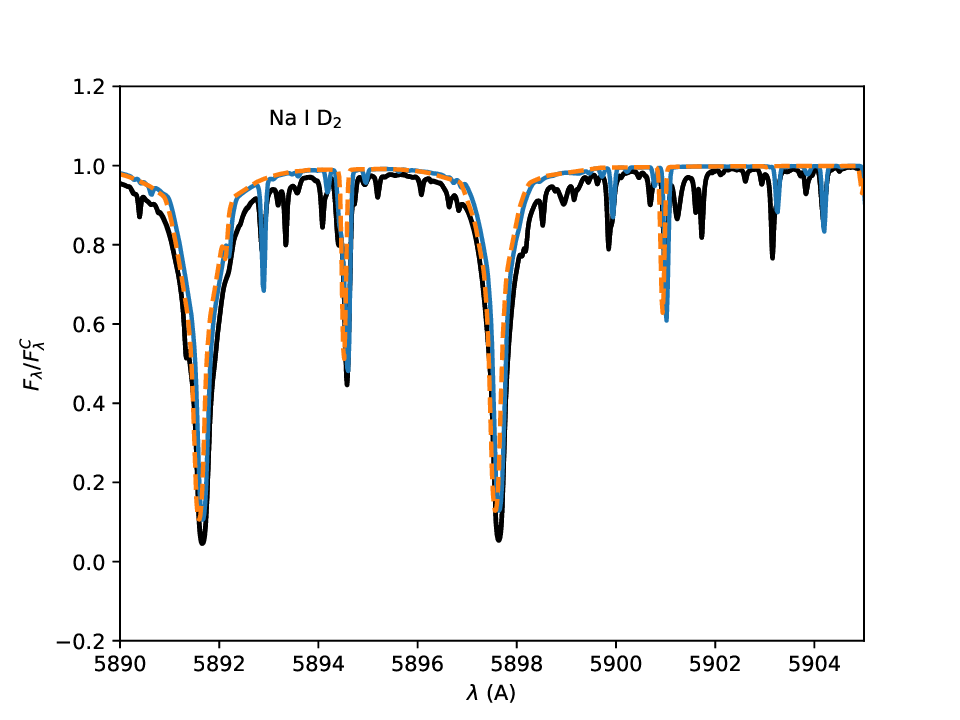}
\caption{The observed and computed high resolution flux spectrum for the Sun in the \ion{Na}{1} $D_{\rm 2}$ 
region.  Observed spectrum of \citet{solarspec} (black line); Synthetic spectra computed with
ChromaStarServer (CSS) with the new VALD line list (solid blue line) and with the previous NIST 
line list (dashed orange line).  The synthetic $F_\lambda$ spectra were broadened with a Gaussian kernel of
$\sigma$ value of 2 km s$^{-1}$.
  \label{SunNormRed}
}
\end{figure}

\begin{figure}
\includegraphics[width=\columnwidth]{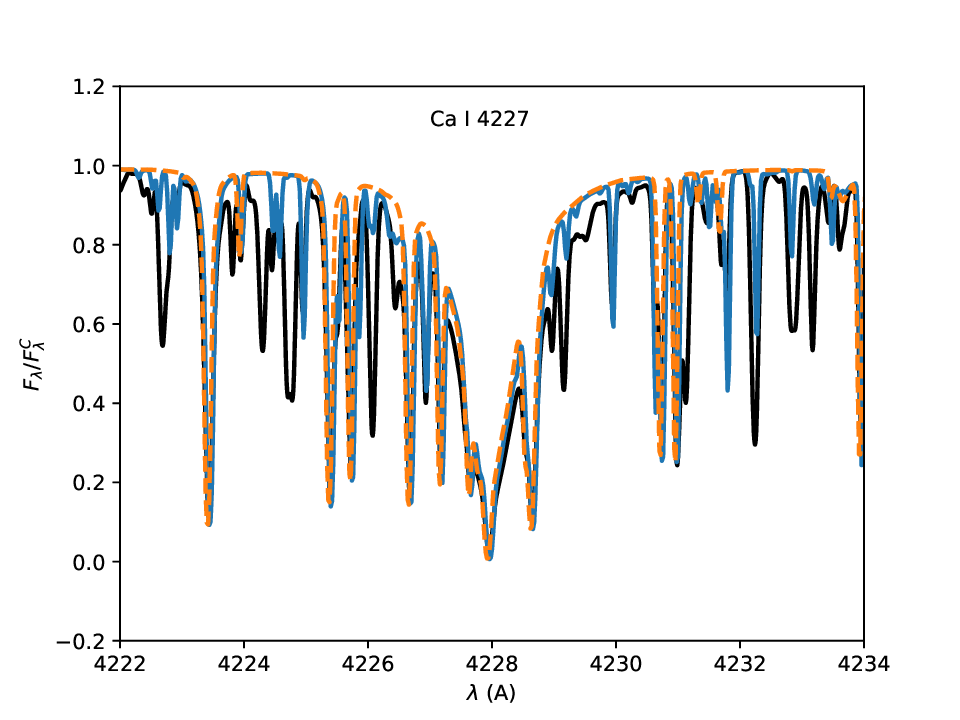}
\caption{Same as Fig. \ref{SunNormRed} except for the \ion{Ca}{1} 4227 line region. 
  \label{SunNormBlue}
}
\end{figure}

\begin{figure}
\includegraphics[width=\columnwidth]{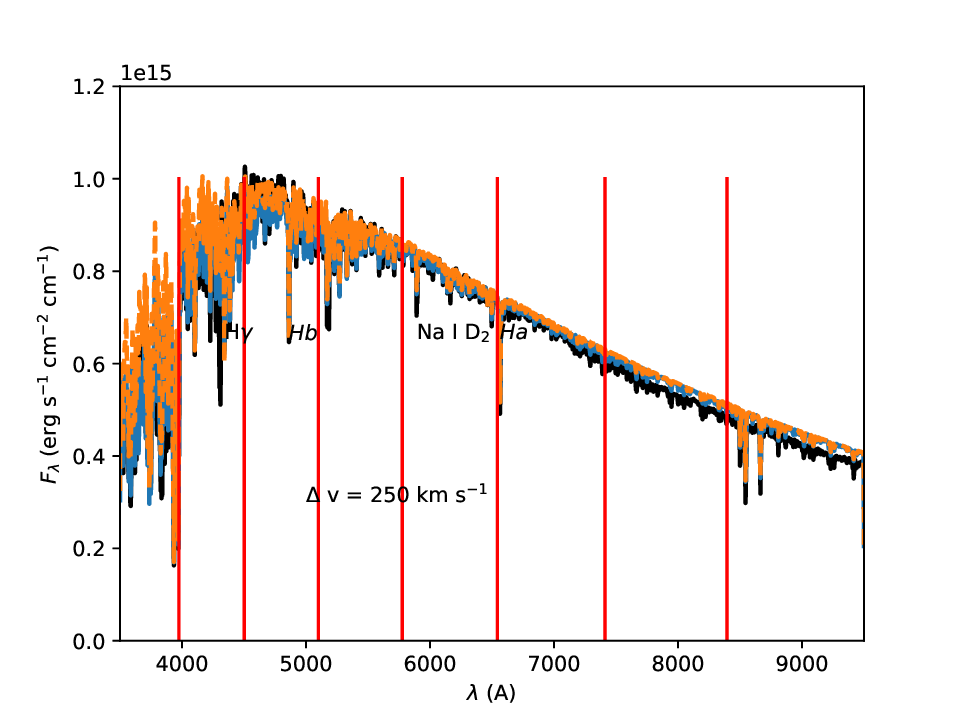}
\caption{The observed and computed spectral energy distribution (SED) for the Sun.  
Observed spectral energy distribution (SED) of \citet{solarsed} (black line); 
Synthetic SEDs computed with
ChromaStarServer (CSS) with the new VALD line list (solid blue line) and with the previous NIST
line list (dashed orange line).  The synthetic $F_\lambda$ spectra were broadened with a Gaussian kernel of
$\sigma$ value of 250 km s$^{-1}$.  The vertical lines indicate the splice points in the spectrum synthesis at
which spectral sub-ranges were automatically scattered to different concurrently running CPUs.  
  \label{SunSed}
}
\end{figure}

\begin{figure}
\includegraphics[width=\columnwidth]{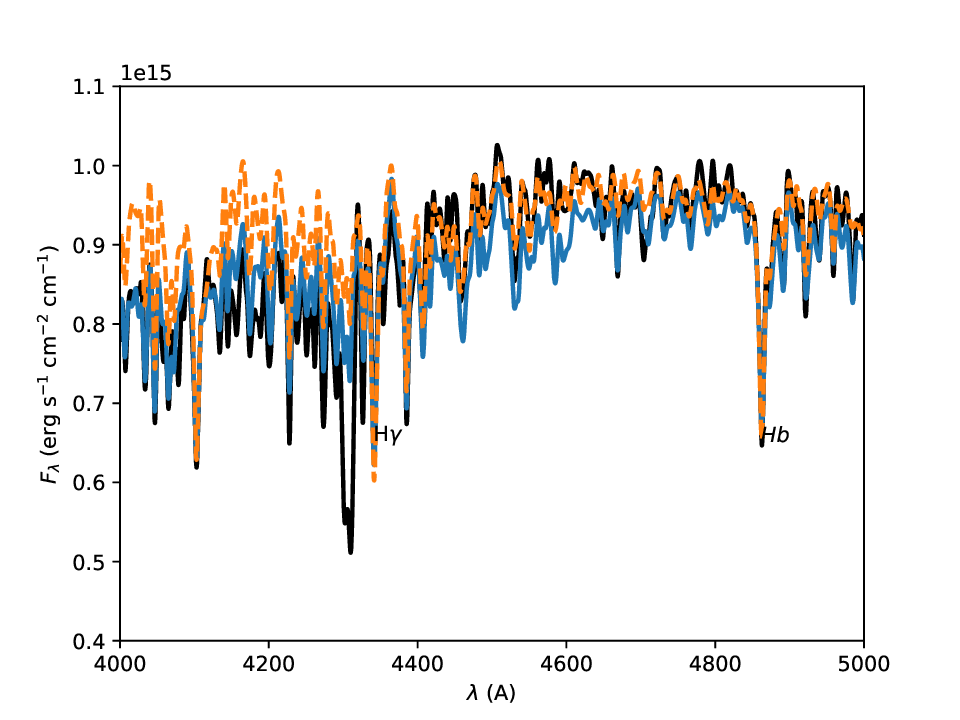}
\caption{Same as Fig. \ref{SunSed} except for the $4000 - 5000$ \AA~ region where the solar SED reaches peak
values.
  \label{SunSedBlue}
}
\end{figure}

Fig. \ref{SunNormRed} shows the continuum-normalized synthetic surface flux spectrum, 
$F_\lambda(\lambda)/F^{\rm C}_\lambda(\lambda)$, for a model of canonical grid input parameters 
close to those of the Sun
 of $(T_{\rm eff}/\log g/[{{\rm A}\over{\rm H}}]) = (5800 K/4.5/0.0)$ as computed with CSS and the NIST and with the VALD
 line lists.  We have used the abundance distribution of \citet{grevs98}, adopted a
 microturbulent velocity dispersion, $\xi_{\rm T}$, of 1 km s$^{-1}$, and a Van der Waals broadening
 enhancement parameter, $\gamma_{\rm VW}$, of $\sim 3$.   
Also shown is the observed solar flux spectrum of \citep{solarflux} of spectral resolution, $R$, of $300\, 000$ 
 in a representative region around the \ion{Na}{1} $D_{\rm 2}$ lines where the spectrum is relatively
 uncrowded.  For this comparison we have smoothed the computed $F_\lambda(\lambda)$ distribution with with a 
Gaussian kernel with a $\sigma$ value of 2.0 km s$^{-1}$.  Fig. \ref{SunNormBlue} shows the same comparison for the 
more crowded \ion{Ca}{1} 4227 line region. 
 Fig. \ref{SunSed} shows the
 spectral energy distribution (SED) of the same model, projected from the Sun's effective surface to the Earth's 
 distance to the Sun along with the observed solar irradiance spectrum of \citep{solarsed}.  For this comparison, both the observed 
 and computed $F_\lambda$ distributions were broadened with a
 Gaussian kernel with a $\sigma$ value of 250 km s$^{-1}$ to allow an assessment of the realism with which the 
 spectral structure is being modelled on the broadband scale.  Fig. \ref{SunSedBlue} shows the same comparison for
 the $\lambda 400$ to $500$ nm region where the Sun's $F_\lambda$ distribution peaks.
 As seen in Figs. \ref{SunSed} and \ref{SunSedBlue}, the new larger line list provides for a more realistically line blanketed 
 SED for GK stars in the $\lambda < 500$ nm range where the SED becomes heavily blanketed and eventually over-blanketed with decreasing 
 $\lambda$ value. 
Fig. \ref{SunSed} also shows the location of the ''Array job'' splice points for the case of the
Sun's spectrum (see section \ref{perf}).  

 \paragraph{$B-V$ color index}

\begin{figure}
\includegraphics[width=\columnwidth]{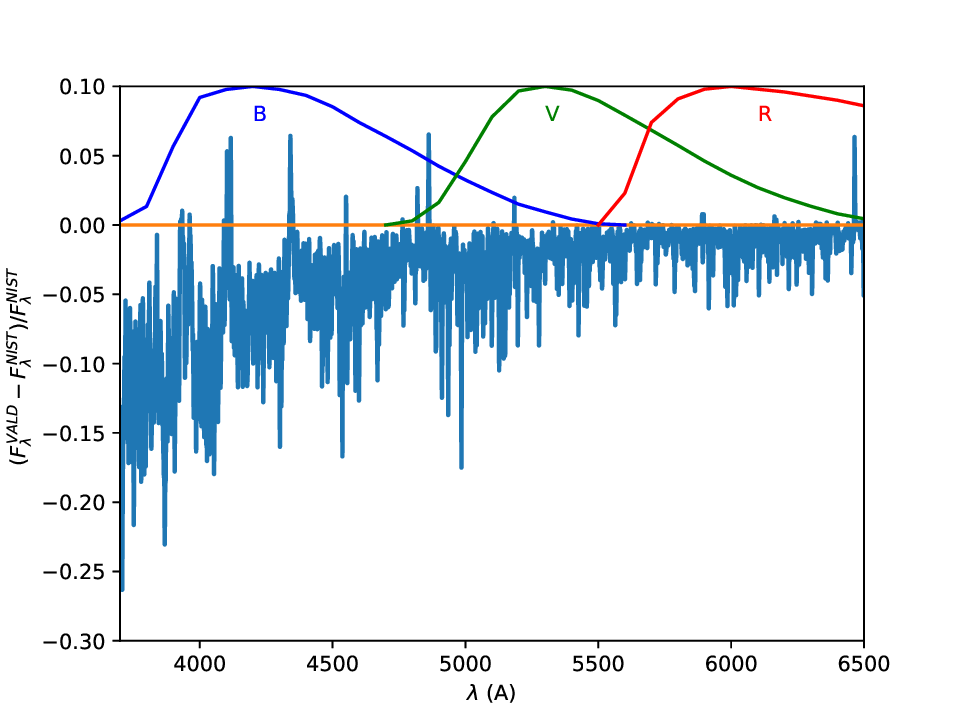}
\caption{The relative difference of synthetic spectra computed with the VALD and NIST line lists, 
$(F^{\rm VALD}_\lambda - F^{\rm NIST}_\lambda)/F^{\rm NIST}_\lambda$,
in the $B-$ and $V-$ spectral regions computed for the Sun with the VALD and NIST line lists, overplotted with the scaled
$B$, $V$, and $R$ transmission curves.  The synthetic $F_\lambda$ spectra were broadened with a Gaussian kernel of 
$\sigma$ value of 100 km s$^{-1}$ before subtraction.
  \label{SunSedDiff}
}
\end{figure}

 The additional line opacity has a small effect on the computed $B-V$ color, {\it decreasing} its value by $\sim 0.015$ mag.    
  Fig. \ref{SunSedDiff} shows the relative difference spectrum 
 $(F^{\rm VALD}_\lambda - F^{\rm NIST}_\lambda)/F^{\rm NIST}_\lambda$ along with the $BVR$ transmission curves
 and reveals that the {\it additional} VALD line opacity has a complicated distribution throughout the $B$ and $V$ bands.
 The synthetic $F_\lambda$ spectra were each broadened by convolution with a Gaussian kernel function with 
 a $\sigma$ value of 100 km s$^{-1}$ before subtraction.
 The calibrated synthetic $B-V$, $V-R$, $V-I$, and $R-I$ color indices in the Johnson-Bessell
 system computed with the smaller NIST line list are $(0.365, 0.717, 1.355, 0.277)$ whereas those computed
 with the new VALD line list are $(0.351, 0.854, 1.530, 0.293)$.  In both cases, the indices are calibrated with a single-point
 correction with a model of Vega computed with the VALD line list and input parameters of
 $(T_{\rm eff}/\log g/[{{\rm A}\over{\rm H}}]) = (9550 K/3.95/-0.5)$ \citep{vega}.  For comparison, the color
 indices computed with our procedure from the observed solar irradiance spectrum of \citet{solarsed} are
 $(0.477, 0.846, 1.408, 0.244)$.

\subsection{Arcturus} 

\begin{figure}
\includegraphics[width=\columnwidth]{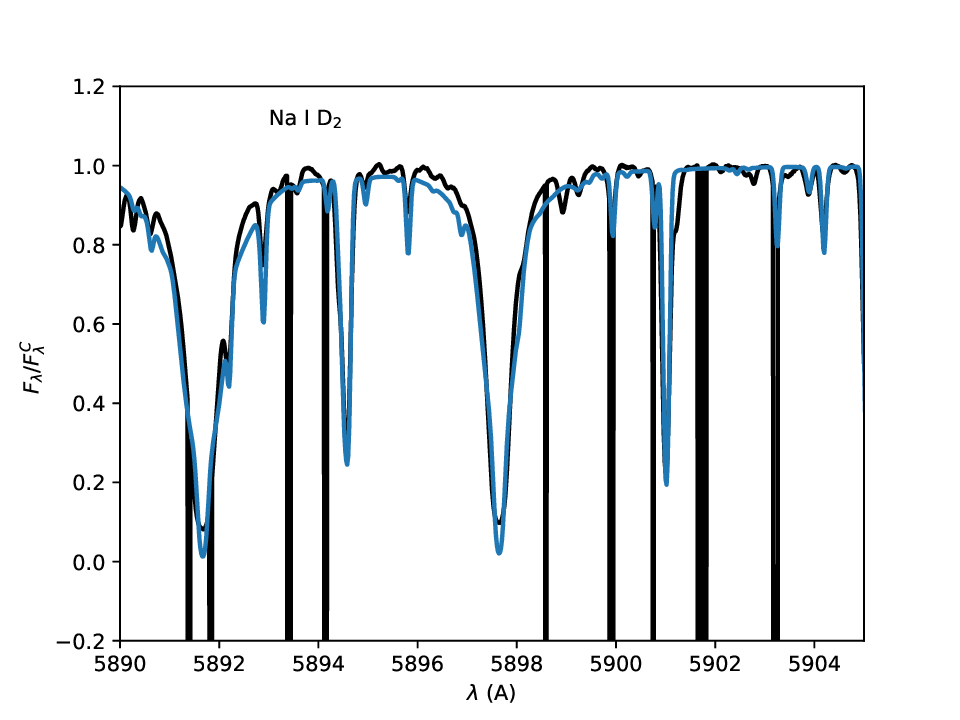}
\caption{Same as Fig. \ref{SunNormRed} except for Arcturus.  The observed spectrum (black)
is that of \citet{hinkle}.  The narrow bands where the observed flux is zero are artifacts in
the published observed spectrum.
  \label{ArcturusNormRed}
}
\end{figure}

\begin{figure}
\includegraphics[width=\columnwidth]{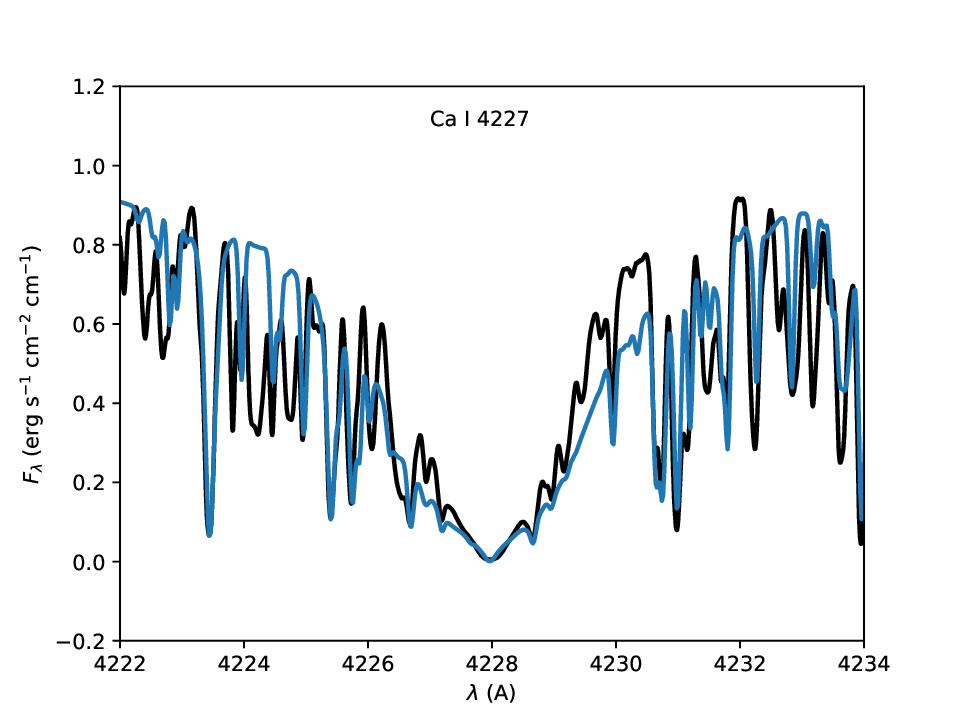}
\caption{Same as Fig. \ref{SunNormBlue} except for Arcturus.  The observed spectrum (black)
is that of \citet{hinkle}. 
  \label{ArcturusNormBlue}
}
\end{figure}

Figs. \ref{ArcturusNormRed} and \ref{ArcturusNormBlue} show the $F_\lambda(\lambda)/F^{\rm C}_\lambda(\lambda)$
distribution for a model of canonical grid point input parameters and scaled solar abundances  
 of $(T_{\rm eff}/\log g/[{{\rm A}\over{\rm H}}]) = (4200 K/1.5/-0.5)$ as computed with the VALD
 line list and smoothed with a Gaussian kernel with a $\sigma$ value of 4.0 km s$^{-1}$ along with 
 the observed flux spectrum of \citet{hinkle} of $R$ value $150\, 000$ 
 in the same \ion{Na}{1} $D_{\rm 2}$ and \ion{Ca}{1} 4227 regions.
 \citet{pdk} carried out a careful spectral analysis of Arcturus with an atomic line list calibrated with a solar model of the 
 Sun's spectrum, and found parameter values of $(T_{\rm eff}/\log g/[{{\rm Fe}\over{\rm H}}]) = (4300 K/1.5/-0.5)$
 with enhanced abundances of the $\alpha$ elements of $+0.3$ dex.  We note that to avoid damped spectral lines with
 wings that are grossly over-broadened, we reduced the $\gamma_{\rm VW}$ value to one 
 ({\it ie.} no enhancement).

\subsection{Vega}	
\begin{figure}
\includegraphics[width=\columnwidth]{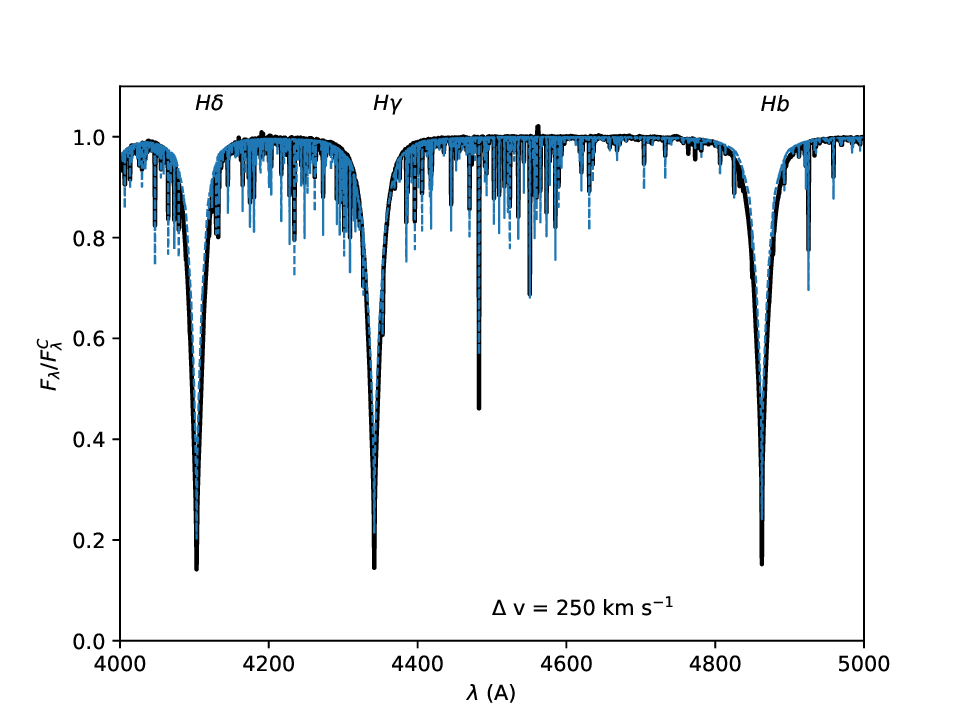}
\caption{The observed and computed high resolution flux spectrum for Vega.  
 Black line: observed spectrum of \citet{vegaspec}; solid blue line: CSS with the VALD line list. 
The synthetic $F_\lambda$ spectrum was broadened with a Gaussian kernel of
$\sigma$ value of 20 km s$^{-1}$.  
%The strong excess observed line opacity in local
%regions of $\lambda < 5700$ \AA~ is caused by telluric absorption.
  \label{VegaNorm}
}
\end{figure}

Fig. \ref{VegaNorm} shows the comparison between the observed high resolution spectrum of Vega of
\citet{vegaspec} and the synthetic spectrum computed with the VALD line list and the input parameters of \citet{vega}  
and the updated \ion{H}{1} broadening throughout the visible band.  As reported in Section \ref{improve} we have
re-tuned the \ion{H}{1} Balmer line linear Stark broadening strengths to match the spectrum of Vega.

%\begin{equation}
%I_{\text{X}^+} = p_{\text{X}^+} \, p_\text{e}/p_\text{X}
%     = \left( \frac{2\pi m_\text{e}kT}{h^2} \right)^{3/2} kT 
%       \left( \frac{2Q_{\text{X}^+}}{Q_\text{X}} \right) \,e^{-\chi_\text{I}/kT}
%\label{IX+}
%\end{equation}

%\begin{multline}
%%\begin{equation}
%\sum_n p_nq_n\sum_k{N_{nk}\over p_{{n}_{k}}}\delta p_{n_k} - 1 \\
%+ {1\over \pe}\sum_np_nq^2_n
% + {p_{\rm Z}^*\over\alpha_{\rm Z}}\sum_m{\alpha_m I_m^+\over (I_{m^+}+\pe)^2 }\delta \pe\\
%  + {1\over\alpha_{\rm Z}}\left(\sum_m {\alpha_m I_m^+}\over {(I_m^+ +\pe)} \right)\delta p_{\rm Z}^* \\
%  = -\sum_n p_n q_n
%  - {p_{\rm Z}^*\over \alpha_{\rm Z}}\sum_m {\alpha_m I_m^+\over (I_m^+}+\pe)
%  + \pe
%\label{charge}
%  \end{multline}
%%\end{equation}

\section{Comparison to Burke-Gaffney Observatory (BGO) spectra}

The Burke-Gaffney Observatory (BGO, lat. $+44\, 37\, 50$, long. $-63\, 34\, 52$) at Saint Mary's 
University (SMU) consists of a 0.6 m (24'')
$f/6.5$ Planewave CDK24 telescope and is  
equipped with an $f/4$ model PF0035 ALPY 600 grism spectrograph from Shelyak Instruments.  We operate the spectrograph
with a slit width of 23 $\mu$m corresponding to a spectral resolving power, $R$, of $\sim 600$, equivalent to
a spectral resolution element of $\Delta v \sim 500$ km s$^{-1}$ at $\lambda\sim 6000$ \AA.  
The camera is a model Atik 314L+ CCD camera with $1391\times 1039$ imaging pixels of size $6.45\times 6.45$ $\mu$m. 
The setup provides a reciprocal linear dispersion, ${\Delta\lambda /\Delta x}$, of $\sim 420$ \AA\, mm$^{-1}$ and a
spectral range, $\Delta\lambda$, of $\sim 3750$ \AA, effectively covering the entire visible band from $\sim 4000$ to $\sim 7000$ \AA~ 
after accounting for edge effects.

\subsection{Observations}

\begin{figure}
\includegraphics[width=\columnwidth]{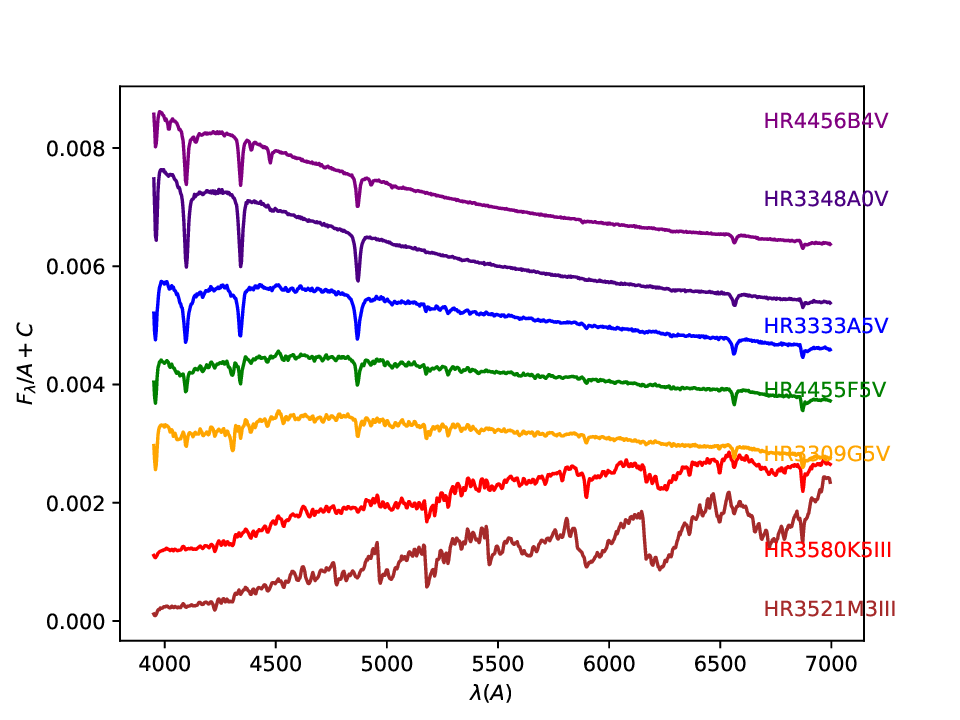}
\caption{Set of seven commissioning spectra for the seven stars listed in Tab. \ref{obslog}
for the ALPY 600 spectrograph at the BGO. 
  \label{DLSequence}
}
\end{figure}

Table \ref{obslog} and Fig. \ref{DLSequence} present a set of commissioning spectra of seven 
bright stars acquired on 
15 April 2021 by the BGO Director
and Astronomy Technician at the time, Mr. David Lane.  The set includes six luminosity class V
stars spanning the range of spectral class from K5 to B4 and one luminosity class III star of
spectral class M3.

\begin{table}
\caption{Commissioning stars observed on 15 April 2021 with the ALPY 600 spectrograph.  All
stellar data are those of \citet{hoffleit}}
\label{obslog}
\begin{tabular}{lrrlr}
\tableline
Designation & $V$ & $B-V$ & Sp. Type & Exp. time (s) \\
\tableline
HR3521  & 6.23 & +1.62  & M3 III & 60 \\
HR3580  & 6.44 & -      & K5 III   & 120\\
HR3309  & 6.32 & +0.62  & G5 V   & 120\\
HR4455  & 5.77 & +0.46  & F5 V   & 120\\
HR3333  & 5.95 & +0.19  & A5 V   & 60\\
HR3348  & 6.18 & -0.03  & A0 V   & 60\\
HR4456  & 5.95 & -0.16  & B4 V   & 120\\
\tableline
\end{tabular}
\end{table}

\subsection{Reduction procedure}

This is the first presentation of BGO spectra in the research literature and in a companion paper 
\citet{bgo} we describe 
in detail our own locally developed BGO post-processing pipeline in the Python programming language for SMU 
students and present a brief summary here.
The pipeline consists of automatic bias, dark current and flat-field corrections performed by software shipped
with the ALPY 600 spectrograph.   That is followed up by 1) Wavelength calibration by fitting a $2^{\rm nd}$-order 
polynomial to the relation ship between pixel columns and the known $\lambda$ value of Ar-Ne lines, 
2) Subtraction of a $0^{\rm th}$-order fit to the remaining residual background to reduce the pedestal signal to zero,
3) Automatic location of the brightness peak in the cross-dispersion profile in sample columns across the 
chip,
4) Generation of a root-$N$ model cross-dispersion weight profile that is centered on the cross-dispersion peak
in each column, 
5) Formation of a $1D$ spectrum by summing columns weighted by the model profile from Step 4),
6) A two-step automatic approximate continuum rectification of the $\sim 4200$ to $6800$ \AA~ region
that is designed for spectra unaffected by emission features or deep TiO bands (spectral classes B to K).

\subsection{Comparison of BGO and CSS spectra}

\begin{figure}
\includegraphics[width=\columnwidth]{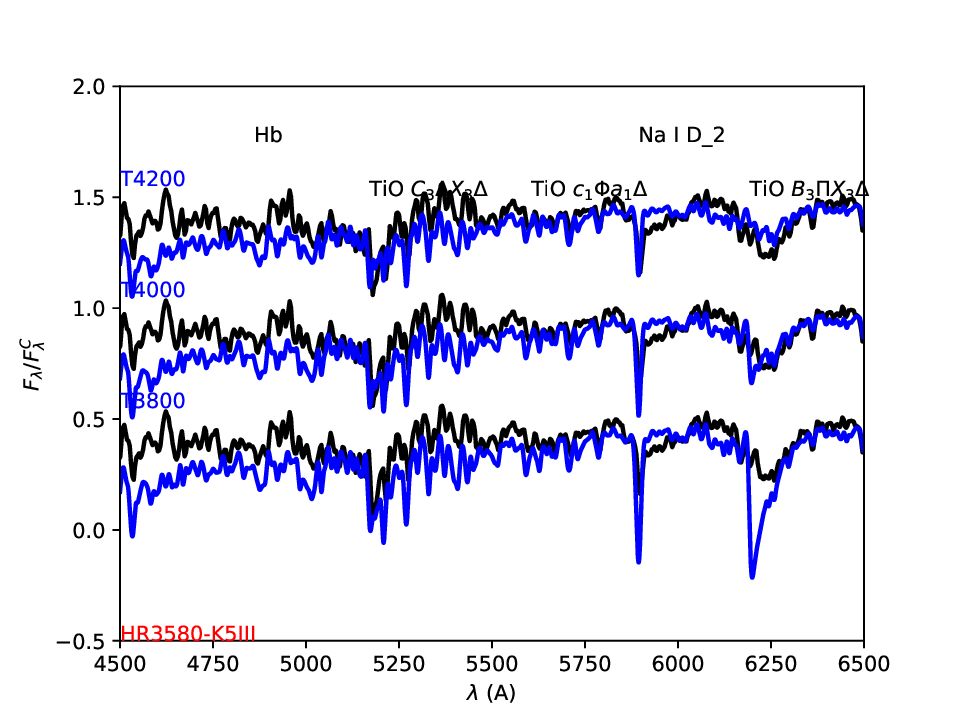}
\caption{HR3580 (K5 III):  Black:  Observed spectrum from our BGO program plotted
with three different vertical offsets.  Blue:
Synthetic spectra for models of $\log g = 2.0$, $[{{\rm A}\over {\rm H}}] = 0.0$, 
and $T_{\rm eff}$ values of 3800, 4000, and 4200 K
in order of increasing vertical offset.
\label{HR3580}
}
\end{figure}

\begin{figure}
\includegraphics[width=\columnwidth]{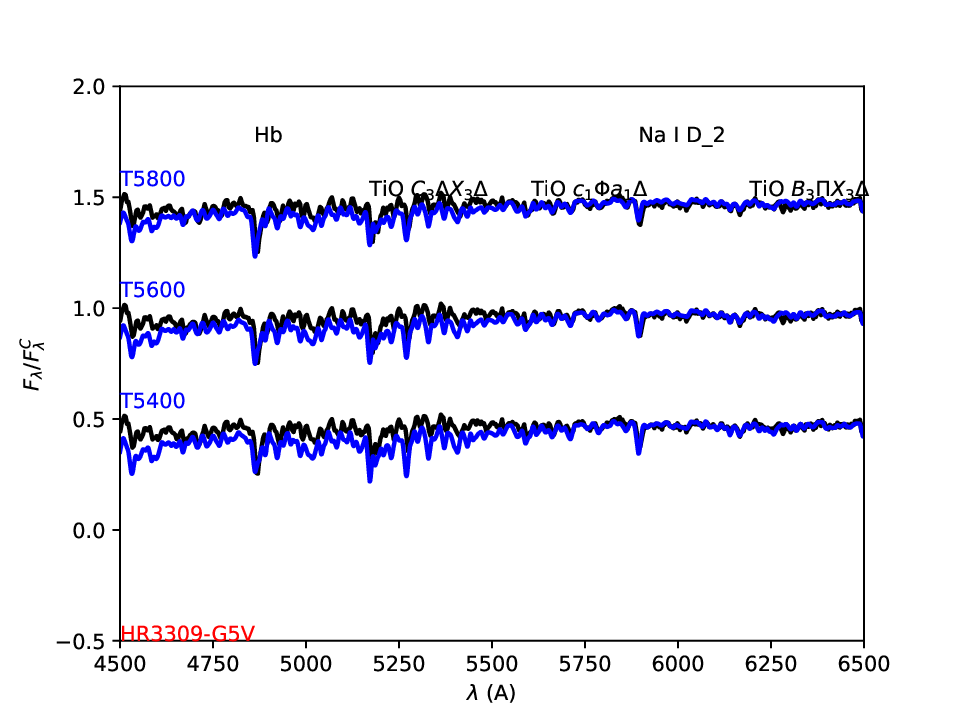}
\caption{Same as Fig. \ref{HR3580} except for star HR3309 (G5 V) and models
of $T_{\rm eff}$ values of 5400, 5600, and 5800 K. 
\label{HR3309}
}
\end{figure}

\begin{figure}
\includegraphics[width=\columnwidth]{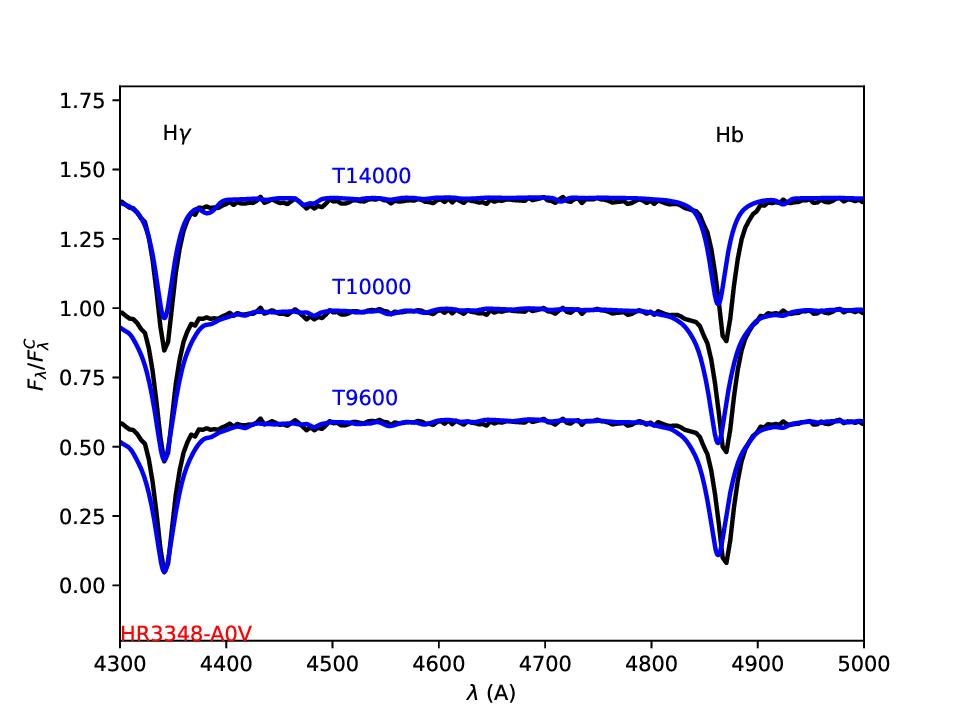}
\caption{Same as Fig. \ref{HR3580} except for star HR3348 (A0 V) and models
of $T_{\rm eff}$ values of 9600, 10000, and 14\, 000 K. 
\label{HR3348}
}
\end{figure}

In Figs. \ref{HR3580}, \ref{HR3309}, and \ref{HR3348} we show the comparison between 
observed spectra from our BGO observing run and synthetic spectra from our model grid 
bracketing the nominal $T_{\rm eff}$ value corresponding to the spectral type listed 
in \citet{hoffleit}.  Nominal $T_{\rm eff}$ values were taken from Appendix G of \cite{appendixG}.
Because we can only achieve an approximate continuum rectification for late type stars 
with high $\Delta\lambda/\Delta x$ broadband data, we do not attempt a quantitative
fit based on minimizing a fitting statistic, but only a perform a visual inspection of the
fit quality.

\paragraph{}
At the low $R$ and high ${{\Delta\lambda}\over{\Delta x}}$ values of these spectra, the main features at which 
we can assess the fit within the $\sim 4200$ to $\sim 6800$ \AA~ rectification range for GK stars are the 
\ion{Na}{1} $D_{\rm 2}$ doublet at $\lambda 5900$ \AA~ and the TiO
C$_{\rm 3}\Delta$-X$_{\rm 3}\Delta$ ($\alpha$ system, $\lambda_{\rm 00}~ 5170$ \AA) and the 
B$_{\rm 3}\Pi$-X$_{\rm 3}\Delta$ ($\gamma$' system, $\lambda_{\rm 00}~ 6193$ \AA) bands.
For A and B stars, at our $R$ and ${{\Delta\lambda}\over{\Delta x}}$ values, the main features at which
we can judge the fit within the rectification range are the \ion{H}{1} 
$\beta$ and $\gamma$ lines.  This is sufficient to allow students to do projects at the 
undergraduate honours level in which they carry out and reduce their own BGO spectroscopy to 
coarsely classify stars to within a few spectral subclasses accuracy.  In the process, they
will gain valuable experience with the procedures of observational and computational stellar spectroscopy
within a Python IDE running on commonplace Windows or Linux computer.

\acknowledgements

This work was made possible by the ACENET research computing consortium (ace-net.ca/) and the Digital Research Alliance of Canada (alliancecan.ca). 
This work has made use of the VALD database, operated at Uppsala University, the Institute of Astronomy RAS in Moscow, and the University of Vienna
(vald.astro.uu.se/).  We gratefully acknowledge the valuable guidance of the BGO technician, Tiffany Fields, and useful discussion with 
Brian Skiff of Lowell Observatory.

\end{document}